\documentclass[aps,prl,twocolumn,amsmath,amssymb,showpacs]{revtex4}
\usepackage{epsfig,float}
\usepackage{graphicx}
\usepackage{dcolumn}
\usepackage{bm}
\usepackage{morefloats}

\begin{document}

\title{The puzzle of anomalously large isospin violations in $\eta(1405/1475)\to 3\pi$}

\author{Jia-Jun Wu$^{1}$}
\author{Xiao-Hai Liu$^{1}$}
\author{Qiang Zhao$^{1,2}$\footnote{zhaoq@ihep.ac.cn}}
\author{Bing-Song Zou$^{1,2}$\footnote{zoubs@ihep.ac.cn} }

\affiliation{1) Institute of High Energy Physics, Chinese Academy of Sciences, Beijing 100049, China\\
2) Theoretical Physics Center for Science Facilities, CAS, Beijing
100049, China}

\begin{abstract}

The BES-III Collaboration recently report the observation of
anomalously large isospin violations in $J/\psi\to
\gamma\eta(1405/1475) \to \gamma \pi^0 f_0(980)\to \gamma +3\pi$,
where the $f_0(980)$ in the $\pi\pi$ invariant mass spectrum appears
to be much narrower ($\sim$ 10 MeV) than the peak width ($\sim$50
MeV) measured in other processes. We show that a mechanism, named as
triangle singularity (TS), can produce a narrow enhancement between
the charged and neutral $K\bar{K}$ thresholds, i.e., $2m_{K^\pm}\sim
2m_{K^0}$. It can also lead to different invariant mass spectra for
$\eta(1405/1475)\to a_0(980)\pi$ and $K\bar{K}^*+c.c.$, which can
possibly explain the long-standing puzzle about the need for two
close states $\eta(1405)$ and $\eta(1475)$ in $\eta\pi\pi$ and
$K\bar{K}\pi$, respectively. The TS could be a key to our
understanding of the nature of $\eta(1405/1475)$ and advance our
knowledge about the mixing between $a_0(980)$ and $f_0(980)$.

\end{abstract}

\date{\today}





\maketitle

The spectrum of excited isoscalar states of $J^{PC}=0^{-+}$, i.e.
radial excitation states of $\eta$ and $\eta^\prime$, is still far
from well-understood. An important and peculiar issue is about the
nature of $\eta(1405)$ and $\eta(1475)$. Although the Particle Data
Group (PDG)~\cite{Nakamura:2010zzi} lists them as two individual
states, there still exist unsolved puzzles in the understanding of
their production and decay properties~\cite{Klempt:2007cp}. In fact,
it is even still controversial whether they are two separated states
or just one state of $0^{-+}$ in different decay
modes~\cite{Klempt:2007cp}.

This situation is greatly improved with the availability of
high-statistic $J/\psi$ and $\psi^\prime$ events from the BES-III
Collaboration. Very recently, the BES-III Collaboration report the
observation of anomalously large isospin violations of the
$\eta(1405/1475)\to 3\pi$ in $J/\psi\to \gamma\eta(1405/1475) \to
\gamma \pi^0 f_0(980)\to \gamma +3\pi$~\cite{besiii-hadron2011}.
What makes this measurement extremely interesting is its involvement
with the $a_0(980)$-$f_0(980)$ mixings, which has also been a
long-standing puzzle in history. In Ref.~\cite{besiii-hadron2011},
it was shown that there is only one enhancement in the vicinity of
1.44 GeV, which corresponds to the $\eta(1405)$ or $\eta(1475)$
state. Moreover, it shows that the $f_0(980)$ signal is only about
10 MeV in width, and the preliminary branching ratio for the
isospin-violating decay $\eta(1405/1475)\to f_0(980) \pi^0$ turns
out to be significant.

At a first glance, this process seems to be complicated by the
correlations between the $\eta(1405/1475)$ ambiguity and
$a_0(980)$-$f_0(980)$ mixings. However, we shall show in this work,
this process would provide a golden opportunity to disentangle the
relation between the $\eta(1405)$ and $\eta(1475)$ signal, and
identify a crucial dynamic mechanism in the isospin-violating decay
of $\eta(1405/1475)\to f_0(980)\pi^0\to 3\pi$, i.e. the triangle
singularity (TS),  where the intermediate on-shell $K\bar{K}^*+c.c.$
pair can exchange an on-shell kaon, and then rescatter to the
isospin violating $f_0(980)\pi^0$.

For this purpose, it is necessary to study the decay of
$\eta(1405/1475)\to a_0(980)\pi\to \eta\pi\pi$ as a correlated
process with $\eta(1405/1475)\to f_0(980)\pi^0\to 3\pi$.
Furthermore, since the TS mechanism is driven by the tree-level
process $\eta(1405/1475)\to K\bar{K}^*+c.c.$, we shall show that
only one $0^{-+}$ isoscalar state is needed here. This ``one state"
assumption is our starting point of this work. To keep the notation
short, we denote it as $\eta(1440)$ in the following analysis.

In this letter, we shall demonstrate that the TS mechanism can lead
to different mass spectra for $\eta(1440)\to K\bar{K}^*+c.c.$ and
$a_0(980)\pi^0$. Namely, due to the TS, the $\eta(1440)$ mass peaks
in $K\bar{K}^*+c.c.$ and $a_0(980)\pi$ would appear differently.
Moreover, anomalously large isospin violations may occur in
$\eta(1440)\to f_0(980) \pi^0$. These features as a consequence of
the TS mechanism could be a natural solution for the long-standing
puzzle about the nature of $\eta(1405/1475)$ in experimental
analyses.

\begin{figure}[htbp] \vspace{-0.cm}
\begin{center}
\includegraphics[width=0.9\columnwidth]{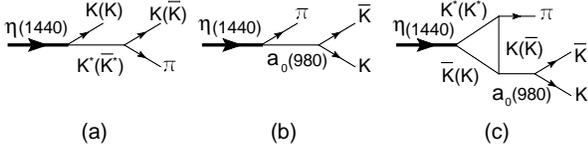}
\caption{The schematic diagrams for $\eta(1440)\to K\bar{K}\pi$.
Diagram (a) and (b) denotes the tree-level transitions, while (c)
illustrates the transition via the TS mechanism. } \label{fig-1}
\end{center}
\end{figure}

\begin{figure}[htbp] \vspace{-0.cm}
\begin{center}
\includegraphics[width=0.8\columnwidth]{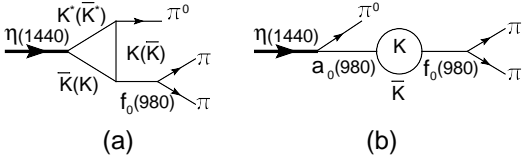}
\caption{The schematic diagrams for $\eta(1440)\to 3\pi$. Diagram
(a) is driven by the TS mechanism, while (b) gives contributions
from the $a_0(980)$-$f_0(980)$ mixing. } \label{fig-2}
\end{center}
\end{figure}

\begin{figure}[htbp] \vspace{-0.cm}
\begin{center}
\includegraphics[width=0.8\columnwidth]{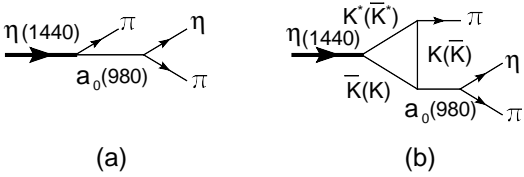}
\caption{The schematic diagram for $\eta(1440)\to \eta\pi\pi$ via
(a) the tree-level $a_0(980)\pi$ production and (b) the TS
mechanism. } \label{fig-3}
\end{center}
\end{figure}

As follows, a coherent investigation of these three transitions,
$\eta(1440)\to K\bar{K}\pi$, $\eta\pi\pi$ and $3\pi$, is presented.
In Figs.~\ref{fig-1}, \ref{fig-2} and \ref{fig-3}, their decays are
illustrated by the schematic Feynman diagrams, respectively. These
transitions are studied in an effective Lagrangian approach with the
vector-vector-pseudoscalar (VVP), vector-pseudoscalar-pseudoscalar
(VPP), and scalar-pseudoscalar-pseudoscalar (SPP) couplings as the
following:
\begin{eqnarray}
\textsl{L}_{V_1V_2P}&=&g_{V_1V_2P}\varepsilon_{\alpha\beta\mu\nu}p^{\alpha}_{V_1}p^{\beta}_{V_2}\psi^{\mu}_{V_1}\psi^{\nu}_{V_2}\psi_{P},\\
\textsl{L}_{VP_1P_2}&=&g_{VP_1P_2}(\psi_{P_1}\partial_{\mu}\psi_{P_2}-\psi_{P_2}\partial_{\mu}\psi_{P_1})\psi^{\mu}_{V},\\
\textsl{L}_{S_1P_2P_3}&=&g_{S_1P_2P_3}\psi_{S_1}\psi_{P_2}\psi_{P_3}
\ ,
\end{eqnarray}
where $\psi_V$, $\psi_P$ and $\psi_S$ stand for the vector,
pseudoscalar, and scalar fields, respectively. The following
relations based on the SU(3) flavor symmetry are implied:
$g_{fK\bar{K}}=g_{fK^+K^-}=-g_{fK^0\bar{K}^0}$,
$g_{aK\bar{K}}=g_{aK^+K^-}=g_{aK^0\bar{K}^0}$,
$\sqrt{2}g_{f\pi\pi}=\sqrt{2}g_{f\pi^0\pi^0}=-g_{f\pi^+\pi^-}$,
$g_{\eta K^* \bar{K}}=g_{\eta K^{*+}K^-}=-g_{\eta
K^{*-}K^+}=-g_{\eta K^{*0}\bar{K}^0}=g_{\eta \bar{K}^{*0}K^0}$,
$g^c_{K^*K\pi}=g_{K^{*+}K^+\pi^0}=-g_{K^{*-}K^-\pi^0}$, or
$g^n_{K^*K\pi}=-g_{K^{*0}K^0\pi^0}=g_{\bar{K}^{*0}\bar{K}^0\pi^0}$.

The coupling constants of $a_{0}(980)$ come from
Ref.\cite{creatball}: $g_{aK\bar{K}}=3.33$ GeV and
$g_{a\eta\pi}=2.45$ GeV. KLOE~\cite{kloe} and BES~\cite{besf} give
different values for $f_{0}(980)$, i.e. KLOE: $g_{fK\bar{K}}=5.92$
GeV, and $g_{f\pi\pi}=2.09$ GeV; BES: $g_{fK\bar{K}}=4.18$ GeV, and
$g_{f\pi\pi}=1.66$ GeV. Fortunately, these two groups of parameters
give almost the same results when the Flatte form of propagator is
applied to $f_{0}(980)$. In other words, the values of the coupling
product $g_{fK\bar{K}}g_{f\pi\pi}G_f$ are almost the same. The
coupling constant $g_{K^*K\pi}$ is calculated from the width of $K^*
\to K\pi$~\cite{Nakamura:2010zzi}, i.e. $g^c_{K^*K\pi}=3.268$ for
the charged channel, and $g^n_{K^*K\pi}=3.208$ for the neutral
channel.

We note that energy-dependent widths are adopted for the $a_0(980)$
and $f_0(980)$ propagators:
\begin{eqnarray}
 G_a&=&{1}/{[s-m^2_{f}+i\sqrt{s}\Gamma_{a}(s)]} \ , \nonumber\\
 G_f&=&{1}/{[s-m^2_{a}+i\sqrt{s}\Gamma_{f}(s)]} \ ,
\end{eqnarray}
where
\begin{eqnarray}
\Gamma_{a}(s)&\equiv &\left\{g^2_{aK\bar{K}}[\rho(K^0,\bar{K}^0)+\rho(K^+, K^-)]\right.\nonumber\\
&&\left.+g^2_{a\pi\eta}\rho(\pi^0, \eta)\right\}/{(16\pi\sqrt{s})}\label{wida},\\
\Gamma_{f}(s)&\equiv &\left\{g^2_{fK\bar{K}}[\rho(K^0, \bar{K}^0)+\rho(K^+, K^-)]\right.\nonumber\\
 &&+\left. g^2_{f\pi\pi}[\rho(\pi^0, \pi^0)+2\rho(\pi^+, \pi^-)]\right\}/{(16\pi\sqrt{s})} \ , \label{widf}
\end{eqnarray}
with $\rho(A,B)\equiv\frac{1}{2s}
[(s-(m_A+m_B)^2)(s-(m_A-m_B)^2)]^{1/2}$.

To compare with the experimental measurement and take into account
the width effects from the $\eta(1440)$ in the typical decay
$J/\psi\to\gamma\eta(1440)\to\gamma ABC$, the following standard
expression is adopted,
\begin{eqnarray}
&&\frac{d \Gamma_{J/\psi \to \gamma\eta(1440)\to \gamma ABC}}{d
\sqrt{s_0}} \nonumber\\
&=&\frac{2s_0}{\pi}\frac{\Gamma_{J/\psi \to \gamma
\eta(1440)}(s_0)\times \Gamma_{\eta(1440) \to
ABC}(s_0)}{(s_0-m^2_{\eta(1440)})^2+\Gamma^2_{\eta(1440)}m^2_{\eta(1440)}}
\end{eqnarray}
where $s_0$ is the four-momentum square of $\eta(1440)$ in the
reaction, and $\Gamma_{\eta(1440)}$ can be $s_0$-dependent:
\begin{eqnarray}
\Gamma_{\eta(1440)}(s_0) =\Gamma_{\eta(1440)\to K\bar{K}^*+c.c. \to
K\bar{K}\pi}(s_0) \ , \label{widtheta}
\end{eqnarray}
which is the tree-level contribution from Fig.~\ref{fig-1}(a). We
note that the contribution of Fig.~\ref{fig-1}(b) via $a_0(980)\pi$
is negligibly small in comparison with Fig.~\ref{fig-1}(a).

Old data for $J/\psi \to \gamma\eta(1405/1475)\to \gamma
K\bar{K}\pi$ were available from  DM2~\cite{dim2}, MARK
III~\cite{mark}, and BES-I~\cite{bes,bes2}. Although there have been
reports of simultaneous observation of two pseudoscalars around 1.44
GeV, as reviewed by the PDG2010~\cite{Nakamura:2010zzi}, the
presence of two states still needs confirmation. Note that most
observations of the $\eta(1405)$ and $\eta(1475)$ are in different
decay channels with different masses. Thus, we first investigate the
invariant mass spectrum of $K\bar{K}\pi$ in $J/\psi\to
\gamma\eta(1440)\to\gamma K\bar{K}\pi$ by fitting the experimental
data~\cite{dim2,mark,bes,bes2}. We then fix the fitted mass
$m_{\eta(1440)}$ and coupling $g_{\eta(1440)K^*\bar{K}}$, and apply
them to $\eta(1440)\to 3\pi$ and $\eta\pi\pi$.

\begin{figure}[htbp] \vspace{-0.cm}
\begin{center}
\includegraphics[width=0.65\columnwidth]{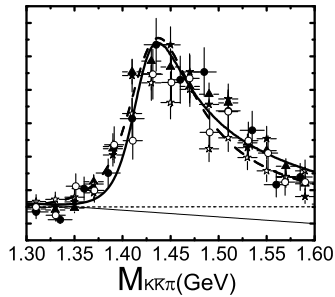}
\caption{The fit for invariant mass spectrum of $K\bar{K}\pi$ with a
constant (thick solid) or energy-dependent (thick dashed line)
width. The solid triangles, solid circles, hollow circles, solid
pentacles, and hollow pentacles are experimental data from MARK III
($K_S^0K^\pm\pi^\mp$)~\cite{mark}, BES-I
($K_S^0K^\pm\pi^\mp$)~\cite{bes}, BES-I ($K^+
K^-\pi^0$)~\cite{bes2}, DM2 ($K_S^0K^\pm\pi^\mp$)~\cite{dim2}, and
DM2 ($K^+ K^-\pi^0$)~\cite{dim2}, respectively. } \label{fit}
\end{center}
\end{figure}

The fit of the invariant mass spectrum of $K \bar{K} \pi$ is shown
in Fig.~\ref{fit}. The thick solid line illustrates the best fit by
a constant width $\Gamma_{\eta(1440)}=67$ MeV at
$m_{\eta(1440)}=1.42$ GeV. The thick dashed line denotes the fit
with an energy-dependent width $\Gamma_{\eta(1440)}(\sqrt{s_0})=166$
MeV at $\sqrt{s_0}=1.42$ GeV and $m_{\eta(1440)}=1.55$ GeV. It is
interesting to see that all those three data sets are fitted well,
and the peak position is about $1.44$ GeV. The background is
estimated by a polynomial form as shown by the thin lines.

In Figs.~\ref{fig-2} and \ref{fig-3}, the $K\bar{K}^*(K)$ loops,
i.e. the TS mechanism, play a dominant role in both $\eta(1440)\to
3\pi$ and $\eta\pi\pi$. For $\eta(1440)\to 3\pi$, the
$a_0(980)$-$f_0(980)$ mixing can also contribute but is relatively
small. We discuss the key features of these transitions as follows:

i) In Fig.~\ref{fig-2}(b), the contribution from the
$a_0(980)$-$f_0(980)$ mixing can be estimated by the $K\bar{K}$
loops, where the incomplete cancellation between the charged and
neutral $K\bar{K}$ loops (due to the mass difference between the
charged and neutral kaons) will lead to non-vanishing transition
matrix element between $a_0(980)$ and $f_0(980)$. The predicted
mixing intensities for $f_0(980)\to a_0(980)$~\cite{af1} and
$a_0(980)\to f_0(980)$~\cite{af2} turn out to be consistent with the
BES-III measurements~\cite{:2010jd} in $J/\psi\to\phi\eta\pi^0$ and
$\chi_{c1}\to 3\pi$, respectively.

Given that the $a_0(980)$-$f_0(980)$ mixing is the exclusive
mechanism in a typical transition of $X\to Y a^0_{0}(980) \to Y
f_0(980) \to Y \pi\pi$, its ratio to the corresponding tree-level
process $X \to Y a^0_{0}(980) \to Y \eta\pi^0$ is about $10^{-2}$.
This allows us to estimate the magnitude of Fig.~\ref{fig-2}(b) in
respect of experimental branching ratio of $3\times10^{-4}$ for
$J/\psi \to \gamma\eta(1405/1475) \to \gamma
\eta\pi^+\pi^-$~\cite{Nakamura:2010zzi}, of which the upper limit is
about $10^{-6}$. This value is nearly one order of magnitude smaller
than the experimental data of $10^{-5}$~\cite{besiii-hadron2011}.
This is a rather direct indication that only the
$a_0(980)$-$f_0(980)$ mixing is not sufficient for accounting for
the new data for $J/\psi \to \gamma\eta(1440) \to \gamma+3\pi$.

ii) The mechanism of Fig.~\ref{fig-2}(a) is very different from the
$a_0$-$f_0$ mixing scheme. It not only produces strong isospin
violations in the subprocess $\eta(1440) \to 3\pi$, but also
enhances such a breaking by the TS mechanism, i.e. within a
particular kinematic region, all those three internal particles
($K\bar{K}^*K$) are allowed to be on-shell.

Note that a full integral of the $K\bar{K}^*(K)$ loop has
ultraviolet divergence. Thus, an empirical form factor generally has
to be introduced in the numerical calculation of the loop
amplitudes, and model-dependence seems to be inevitable. However, it
should be recognized that the absorptive part of the loop integral
is rather model-independent here. Because of the presence of the
on-shell kinematics for all those internal particles, i.e. the TS
mechanism, the absorptive part of the loop integral can be
calculated directly in the on-shell approximation.

\begin{figure}[htbp] 
\begin{center}
\includegraphics[width=0.49\columnwidth]{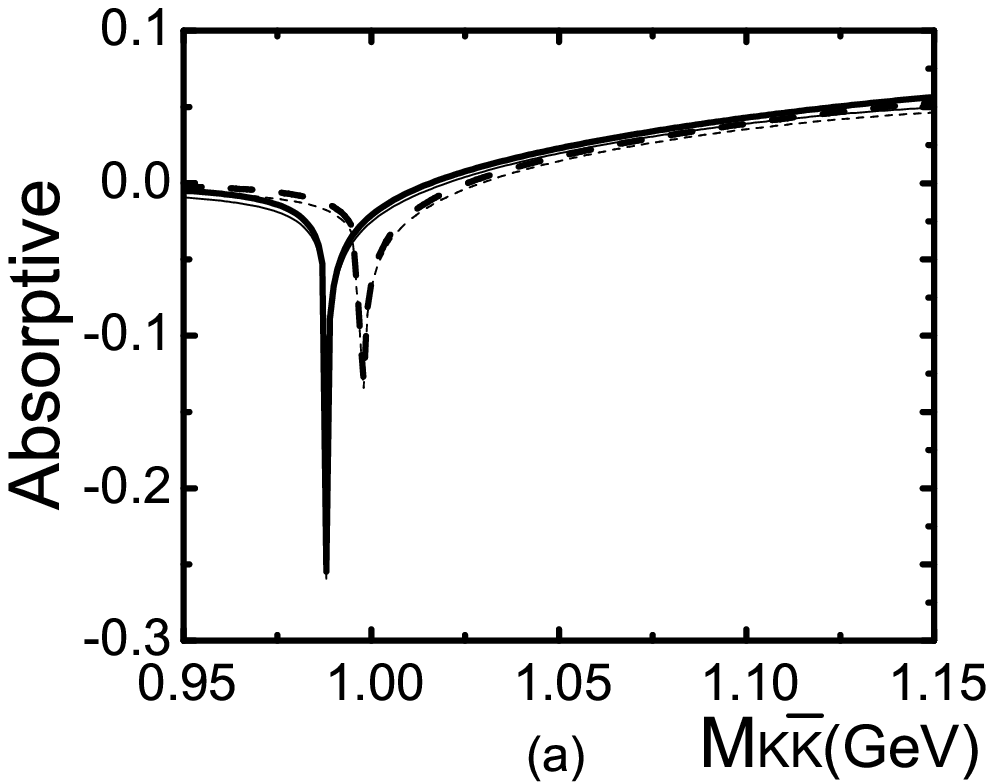}
\includegraphics[width=0.49\columnwidth]{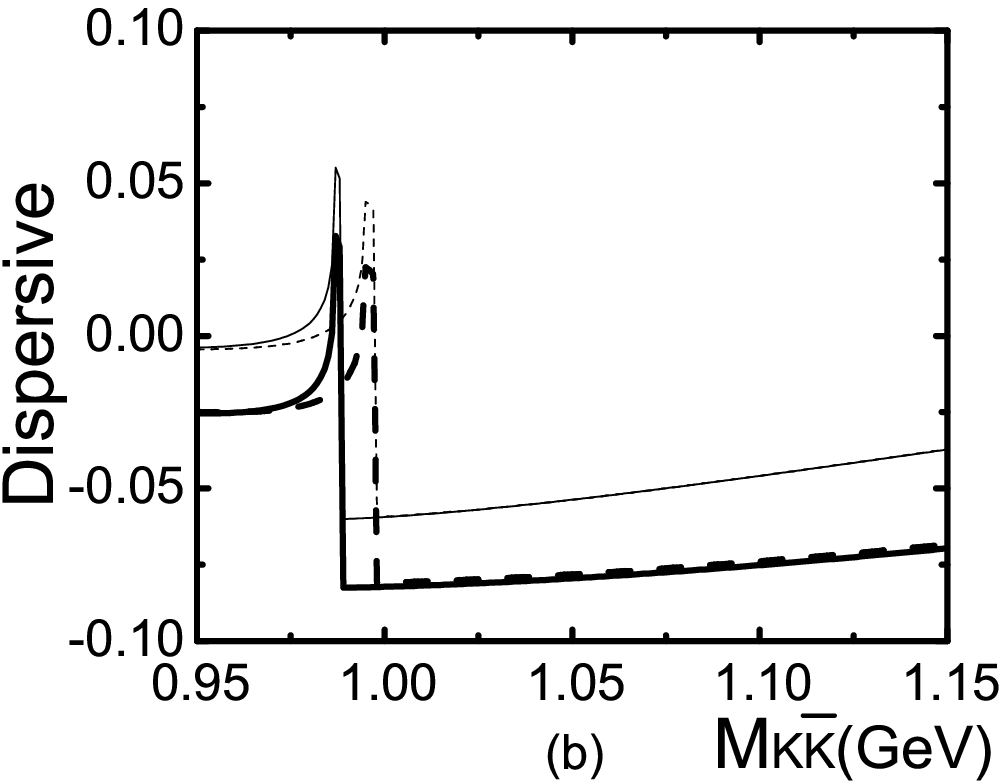}
\caption{The absorptive (a) and dispersive part (b) of the loop
amplitude versus the invariant mass of $K \bar{K}$ at
$\sqrt{s_0}=1.42$ GeV for the $\eta(1440)$, with solid and broken
curves for the charged and neutral $K\bar K^*$ loops, respectively.
} \label{loop3fig}
\end{center}
\end{figure}

In Fig.~\ref{loop3fig}, the absorptive and dispersive part of the
loop integrals are calculated using different cut-off schemes. We
skip the details of form factors adopted here, but just point out
that the absorptive parts of the charged and neutral loops are
predominantly driven by the TS, and insensitive to the form factors.
In Fig.~\ref{loop3fig}(a) the peaking positions of the absorptive
amplitudes for the charged and neutral loops in the $K\bar{K}$
invariant mass spectrum are given by the thresholds of
$K^0\bar{K}^0$ and $K^+ K^-$ due to the TS condition. In contrast,
we find a sensitivity of the dispersive part to the form factors as
shown by Fig.~\ref{loop3fig}(b). Interestingly, it shows that the
cancellation between the charged and neutral amplitude is still
insensitive to the form factor. This indicates a model-independent
feature of the isospin-violating amplitude via the TS mechanism.
Moreover, we confirm the dominance of the TS mechanism
(Fig.~\ref{fig-2}(b)) in $J/\psi\to\gamma\eta(1440)\to\gamma +3\pi$.
For the isospin-conserved transition,
$J/\psi\to\gamma\eta(1440)\to\gamma\eta\pi\pi$, the dispersive part
becomes strongly model-dependent since it is given by the sum of the
charged and neutral amplitudes.

\begin{figure}[htbp] \vspace{-0.cm}
\begin{center}
\includegraphics[width=0.65\columnwidth]{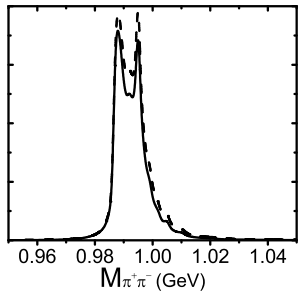}
\caption{The invariant mass spectrum of $\pi^+\pi^-$ from the TS
mechanism in Fig.~\ref{fig-2}(b). The solid line is given by
calculations with a constant width for $\eta(1440)$, while the
dashed line is with an energy-dependent width.} \label{fig-6}
\end{center}
\end{figure}

iii) Because of the dominance of TS mechanism in Figs.~\ref{fig-2}
and \ref{fig-3}, it predicts a narrow peak around the $f_0(980)$ in
the invariant mass spectrum of $\pi^+\pi^-$ due to the cancellation
between the charged and neutral $K\bar{K}^*(K)$ loops, while the
$a_0(980)$ in the $\eta\pi^0$ invariant mass spectrum is not
necessarily narrow. Note that the $f_0(980)$ has a peak width about
50 MeV~\cite{Nakamura:2010zzi}, a much narrower structure ($\sim 10$
MeV) around $f_0(980)$ as demonstrated in Fig.~\ref{fig-6} is an
indisputable signature for the TS mechanism.

\begin{figure}[htbp] \vspace{-0.cm}
\begin{center}
\includegraphics[width=0.49\columnwidth]{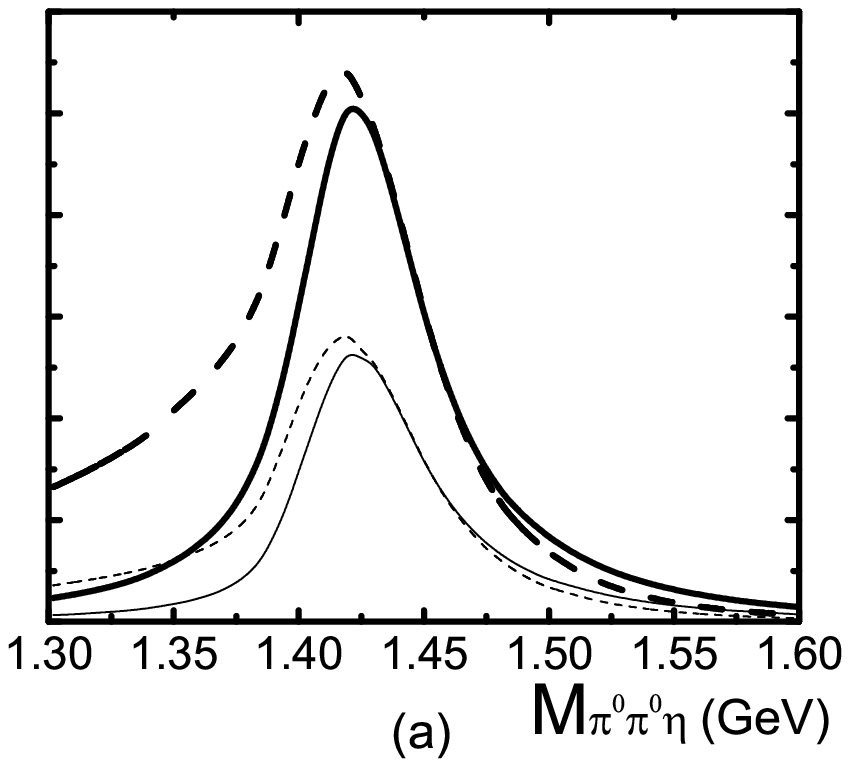}
\includegraphics[width=0.49\columnwidth]{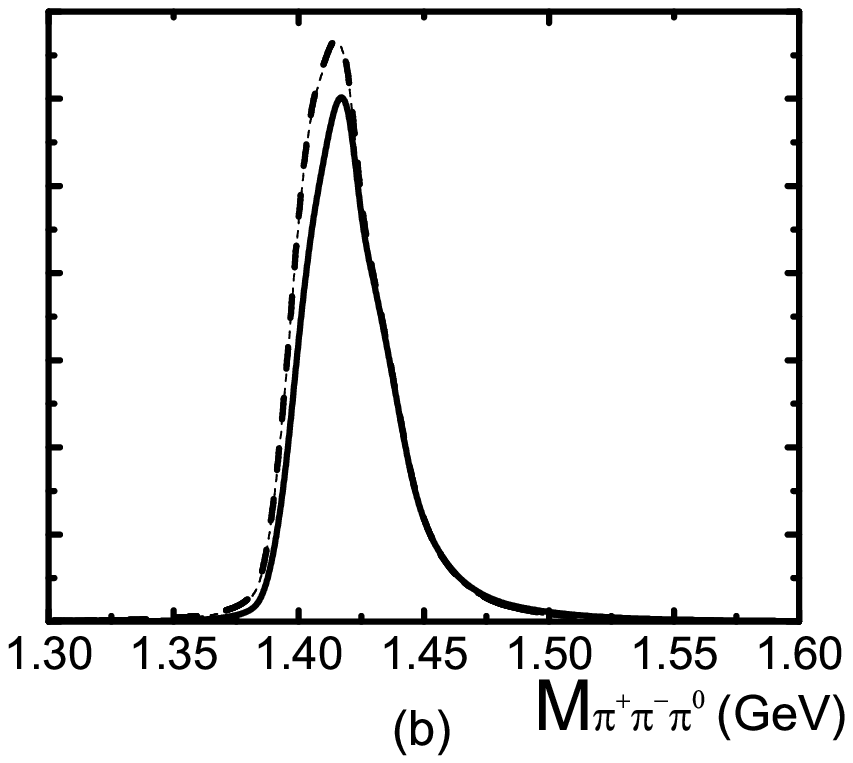}
\caption{The invariant mass spectra of (a) $\eta\pi^0\pi^0$ and (b)
$\pi^+\pi^-\pi^0$. The solid lines are constant-width calculations
for $\eta(1440)$, while the dashed lines are energy-dependent-width
calculations. The thick and thin lines in (a) and (b) are results
using different cut-off energies in the form factor, respectively. }
\label{fig-7}
\end{center}
\end{figure}

iv) In Fig.~\ref{fig-7}, the invariant mass spectra of
$\eta\pi^0\pi^0$ and $\pi^+\pi^-\pi^0$ are plotted with the same
pole mass (1.42 GeV) and width as in $\eta(1440)\to K\bar{K}\pi$.
Interestingly, the peak position is obviously shifted by the TS
terms, and appears to be about $1.415$ GeV. Recall that in
Fig.~\ref{fit} the peak position is around 1.44 GeV. It shows that
the TS mechanism should be responsible for the observed mass
difference for $\eta(1405/1475)$ in the $\eta\pi^0\pi^0$ and
$K\bar{K}\pi$ spectrum. It is worth mentioning that the invariant
mass spectrum of $\pi^+\pi^-\pi^0$ is totally independent of the
cut-off energy of the form factor since the difference of charged
and neutral loop contributions is basically independent of the
cut-off parameter. This TS mechanism also gives correctly the
relative production rates for $\eta(1440)\to K\bar K\pi$,
$\eta\pi\pi$ and $\pi^+\pi^-\pi^0$.

To summarize, we have studied the $J/\psi \to \gamma\eta(1405/1475)$
decays into three different final states: $K\bar{K}\pi$,
$\pi^+\pi^-\pi^0$ and $\eta\pi^0\pi^0$. By the assumption that only
one state of $0^{-+}$ is present around 1.44 GeV, i.e. $\eta(1440)$,
our results are in good agreement with the experimental data. The
dynamic reason is because of the TS mechanism, which allows all the
internal particles to be on-shell in this particular kinematic
region. This mechanism turns out to be much more dominant than the
$a_0$-$f_0$ mixing term, and can thus lead to the anomalously large
isospin violations in $\eta(1440) \to \pi^+\pi^-\pi^0$.
Nevertheless, the peak positions of the $\eta(1440)$ in the
invariant mass spectra of $K\bar{K}\pi$, $\pi^+\pi^-\pi^0$ and
$\eta\pi^0\pi^0$ are shifted by the TS amplitudes and appear to have
different lineshapes. This phenomenon can possibly explain the
puzzling presence of two states $\eta(1405)$ and $\eta(1475)$ in the
previous data analysis, and improve our understanding of the
isoscalar pseudoscalar spectrum~\cite{Yu:2011ta}. This clarification
that only one state $\eta(1440)$ is eventually needed in this energy
region is nontrivial since it could also provide important insights
into the lightest pseudoscalar glueball candidate, which could be
much heavier than $\eta(1440)$ as predicted by the lattice
QCD~\cite{Bali:1993fb}. Further relevant issues can be studied by
BES-III experiment in the near future.

Useful discussions with X.-Y. Shen, S. Jin, H.-B. Li, and B.-J. Liu
are acknowledged. This work is supported, in part, by the National
Natural Science Foundation of China (Grant Nos. 11035006, 10875133,
10821063), the Chinese Academy of Sciences (KJCX2-EW-N01), and the
Ministry of Science and Technology of China (2009CB825200).

\end{document}